\documentclass[3p]{elsarticle}
\makeatletter
\def\ps@pprintTitle{%
 \let\@oddhead\@empty
 \let\@evenhead\@empty
 \def\@oddfoot{}%
 \let\@evenfoot\@oddfoot}
\makeatother
\usepackage{graphicx}
\usepackage{amsmath}
\usepackage{tikz}
\usetikzlibrary{circuits.logic.US} 
\usepackage{mathtools}
\usepackage[font=footnotesize]{caption}
\usepackage{float}
\usepackage{lineno}

\begin{document}
\bibliographystyle{unsrt}

\title{HiCal 2: An Instrument Designed for Calibration of the ANITA Experiment and for Antarctic Surface Reflectivity Measurements}

\author[KU]{S.~Prohira}
\author[KU,MEPhI]{A.~Novikov}
\author[KU,MEPhI]{D.~Z.~Besson}
\author[KU]{K.~Ratzlaff}
\author[KU]{J.~Stockham}
\author[KU]{M.~Stockham}
\author[UD]{J.~M.~Clem}
\author[KU]{R.~Young}

\author[UH]{P.~W.~Gorham}


\author[OSU,CCAP]{P.~Allison}

\author[OSU]{O.~Banerjee}

\author[UCL]{L.~Batten}

\author[OSU,CCAP]{J.~J.~Beatty}

\author[JPL]{K.~Belov}


\author[WashU]{W.~R.~Binns}

\author[WashU]{V.~Bugaev}

\author[UD]{P.~Cao}

\author[NTU]{C.~Chen}

\author[NTU]{P.~Chen}


\author[OSU,CCAP]{A.~Connolly}

\author[UCL]{L.~Cremonesi}

\author[OSU]{B.~Dailey}


\author[Chicago]{C.~Deaconu}


\author[UCLA]{P.~F.~Dowkontt}



\author[UH]{B.~D.~Fox}


\author[OSU]{J.~Gordon}

\author[SLAC]{C.~Hast}


\author[UH]{B.~Hill}


\author[OSU]{R.~Hupe}

\author[WashU]{M.~H.~Israel}



\author[UH]{J.~Kowalski}

\author[UCLA]{J.~Lam}

\author[UH]{J.~G.~Learned}

\author[JPL]{K.~M.~Liewer}

\author[NTU]{T.C. Liu}

\author[Chicago]{A.~Ludwig}



\author[UH]{S.~Matsuno}


\author[UH]{C.~Miki}


\author[UCL]{M.~Mottram}

\author[UD]{K.~Mulrey}



\author[NTU]{J.~Nam}

\author[UCL]{R.~J.~Nichol}


\author[Chicago]{E.~Oberla}




\author[WashU]{B.~F.~Rauch}


\author[UH]{J.~Roberts}


\author[JPL]{A.~Romero-Wolf}

\author[UH]{B.~Rotter}


\author[UH]{J.~Russell}


\author[UCLA]{D.~Saltzberg}

\author[UH]{H.~Schoorlemmer}

\author[UD]{D.~Seckel}


\author[OSU]{S.~Stafford}



\author[UCLA]{B.~Strutt}

\author[UH]{K.~Tatem}

\author[UH]{G.~S.~Varner}

\author[Chicago]{A.~G.~Vieregg}

\author[CalPoly]{S.~A.~Wissel}

\author[UCLA]{F.~Wu}


\address[UCLA]{Dept.~of Physics and Astronomy, Univ.~of California, Los Angeles, Los Angeles, CA 90095.}
\address[OSU]{Dept.~of Physics, Ohio State Univ., Columbus, OH 43210.}
\address[UH]{Dept.~of Physics and Astronomy, Univ. of Hawaii, Manoa, HI 96822.}
\address[NTU]{Dept.~of Physics, Grad. Inst. of Astrophys.,\& Leung Center for Cosmology and Particle Astrophysics, National Taiwan University, Taipei, Taiwan.}
\address[KU]{Dept.~of Physics and Astronomy, Univ. of Kansas, Lawrence, KS 66045.}
\address[WashU]{Dept.~of Physics, Washington Univ. in St. Louis, MO 63130.}
\address[SLAC]{SLAC National Accelerator Laboratory, Menlo Park, CA, 94025.}
\address[UD]{Dept.~of Physics, Univ. of Delaware, Newark, DE 19716.}
\address[UCL]{Dept.~of Physics and Astronomy, University College London, London, United Kingdom.}
\address[JPL]{Jet Propulsion Laboratory, Pasadena, CA 91109.}
\address[CCAP]{Center for Cosmology and Particle Astrophysics, Ohio State Univ., Columbus, OH 43210.}
\address[Chicago]{Dept.~of Physics, Enrico Fermi Institute, Kavli Institute for Cosmological Physics, Univ. of Chicago , Chicago IL 60637.}
\address[CalPoly]{Dept.~of Physics, California Polytechnic State Univ., San Luis Obispo, CA 93407.}
\address[MEPhI]{National Research Nuclear University, Moscow Engineering Physics Institute, 31 Kashirskoye Highway, Russia 115409}
%


\begin{abstract}
The NASA supported High-Altitude Calibration (HiCal)-2 instrument flew as a companion balloon to the ANITA-4 experiment in December 2016. Based on a high-voltage (HV) discharge pulser producing radio-frequency (RF) calibration pulses, HiCal-2 comprised two payloads, which flew for a combined 18 days, covering 1.5 revolutions of the Antarctic continent. ANITA-4 captured over 10,000 pulses from HiCal-2, both direct and reflected from the surface, at distances varying from 100--700~km, providing a large dataset for surface reflectivity measurements. Herein we present details on the design, construction and performance of HiCal-2.

\textit{Keywords: }Radio; Antarctica; Ice properties; High-voltage; Pulse; Pulser
\end{abstract}
\maketitle
\section{Introduction: The HiCal project and ANITA}
\begin{figure}[H]
\centering
\includegraphics[width=.9\textwidth]{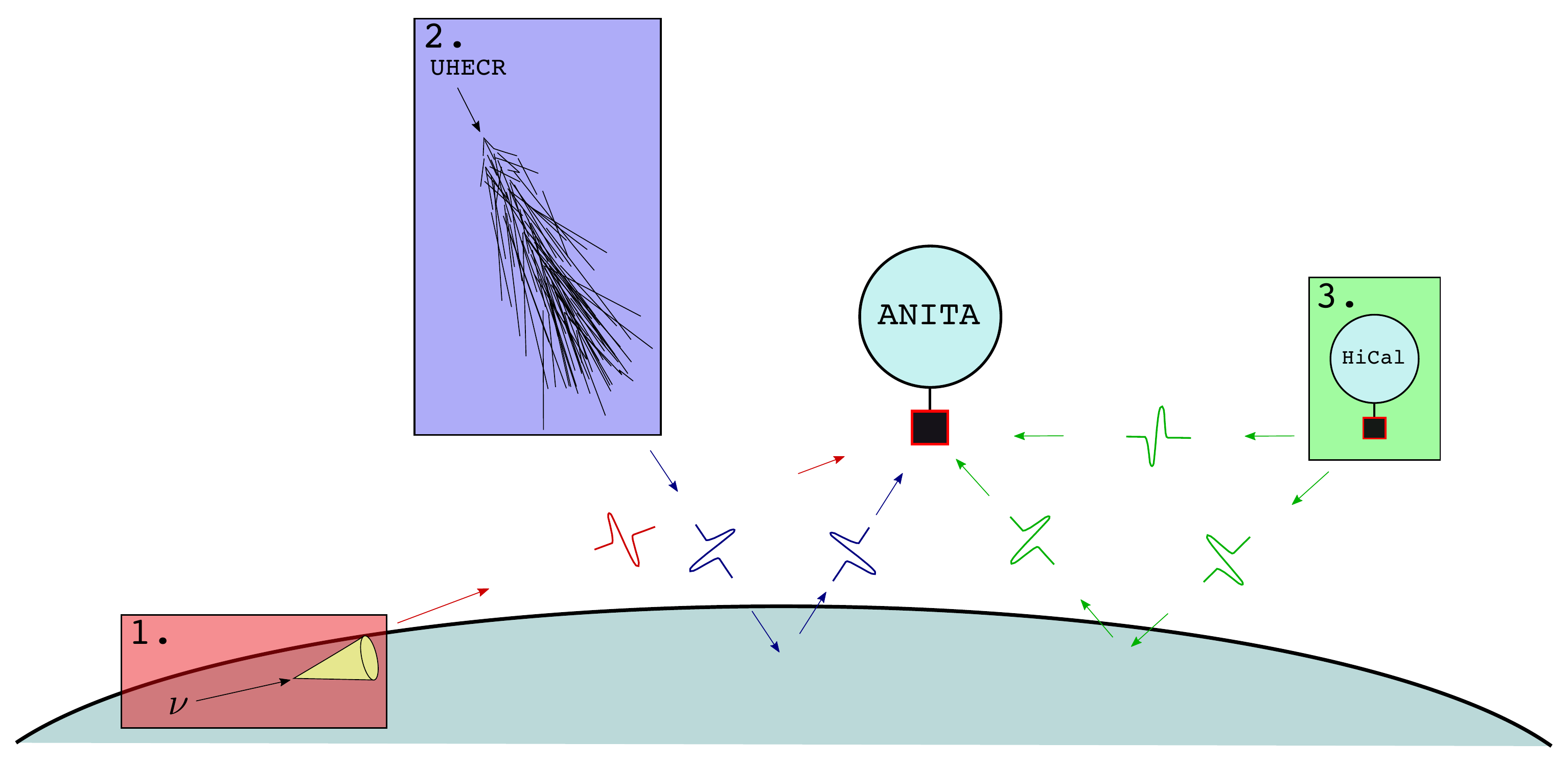}
\caption{The various signals searched for by ANITA. 1) Neutrino interactions with ice molecules create cones of Askaryan radiation, which upon exit from the ice, are detected by ANITA as short impulsive transients. 2) UHECR showers produce radio by a mix of the Askaryan effect and geomagnetic deflection, and are generally down-going. RF signals from downgoing UHECR are detected after being reflected up off the ice (and occasionally directly for slightly upward going showers near the horizon). 3) HiCal produces an impulsive RF signal to mimic the UHECR signal. ANITA receives the HiCal signal both directly and also reflected from the ice. }\label{anita}
\end{figure}

The Antarctic Impulsive Transient Antenna (ANITA)~\cite{anita} is a balloon-borne antenna array instrument that searches for Askaryan radio emissions from interactions of ultra-high-energy cosmic ray neutrinos with the Antarctic ice~\cite{askaryan_orig}. The ANITA instrument is also sensitive to ultra-high-energy cosmic ray (UHECR) radio signals \cite{anita_cr}, which are typically detected after the radio signal from the down-going shower is reflected up off of the ice surface. One important experimental uncertainty is the extent to which the surface roughness affects the received signal amplitude and spectrum of cosmic rays. The HiCal instrument is a balloon-borne high-voltage (HV) pulser that either leads or follows the ANITA payload on a separate balloon and periodically emits impulsive broadband RF signals. These radio pulses can be received by ANITA twice, both direct and reflected from the Antarctic surface, simultaneously calibrating the ANITA instrument and providing measurements of the surface reflectivity at various incidence angles. The various signals of interest to which ANITA is sensitive are shown diagrammatically in Figure~\ref{anita}. 

HiCal-1 flew successfully in December 2014 as a companion balloon to ANITA\-3~\cite{hical1}. In this paper, we discuss an improved system, HiCal-2, that accompanied ANITA-4 on its circumpolar journey in December 2016. 

A theoretical description of the surface reflectivity problem is given in~\cite{harm}, which quantifies the extent to which surface roughness can affect the reconstructed primary UHECR energy. To test this model, one can analyze the ratio of reflected vs. direct signal amplitudes of various signals as seen by the ANITA instrument. An analysis of surface reflectivity using the sun during the ANITA-2 flight is given in~\cite{a2solar}, and an analysis of satellite, solar, and HiCal-1 data during the ANITA-3 flight is given in~\cite{hical1}. The HiCal-1 data represented the first transient signals to be analyzed in such a way, but unfortunately the flight path only allowed for measurements at very large separation distances, indicating some discrepancy between model calculation and data at such glancing surface incidence angles. Specifically, the reflected vs. direct power ratio of pulses seen by ANITA-3 exceeded predictions in the elevation angle (complement of the zenith angle) range of 3-6 degrees~\cite{hical1}. HiCal-2 was flown to provide considerably improved statistics over a wider range of incidence angles, and investigate further the apparent disagreement between model and HiCal-1 data.

\subsection{HiCal-2 subsystems}
HiCal-2 consisted of the 4 main sub-systems shown in Figure~\ref{hical}: \begin{enumerate}
\item The HV pulse generator and antenna, inside of a 1 atmosphere Pressure Vessel (PV).
\item Communication, telemetry, and GPS information subsystems provided by NASA’s Columbia Scientific Balloon Facility (CSBF) to and from the instrument, as well as the battery power supply.
\item The Azimuth and Time-Stamp Apparatus (ATSA), for heading and environmental data.
\item The HiCal system board, which timestamped the outgoing pulses and combined PV pressure and ATSA data into packets to be telemetered.
\end{enumerate} 

Collectively, these systems were constrained by a 5~kg total weight limit, which proved to be an engineering challenge. Design choices had to be made to maximize performance of the instrument, while at the same time minimizing weight. 

\begin{figure}[H]
\centering
\includegraphics[width=.5\textwidth]{hicalannotated}
\caption{The HiCal-2 instrument. For scale reference, the pressure vessel is $\sim$0.61~m from end to end.}\label{hical}
\end{figure}

\section{High-Voltage Pulse Generation} 
\subsection{Piezoelectricity}
\begin{figure}[H]
\centering
\includegraphics[width=.5\textwidth]{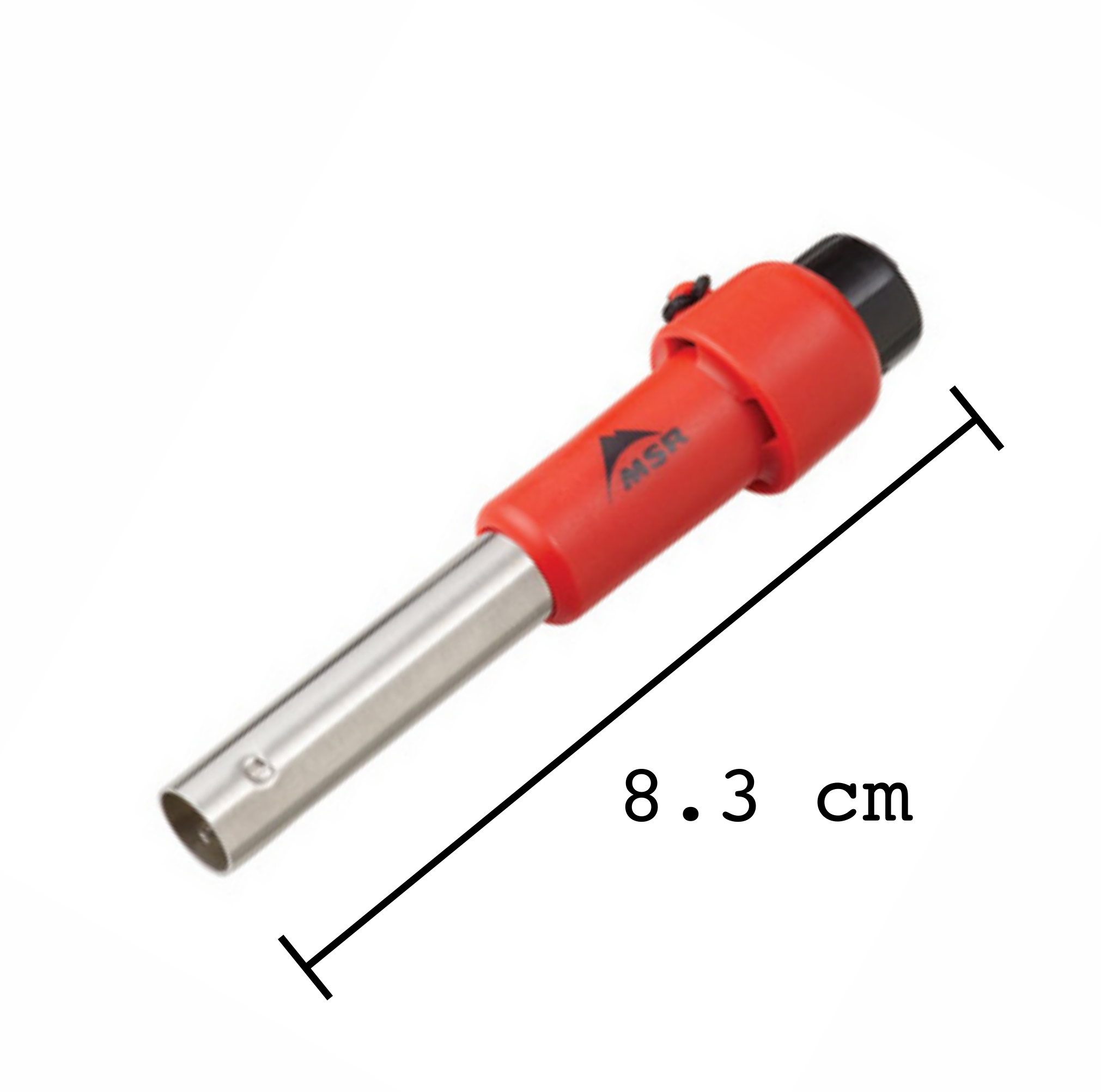}
\caption{The MSR camp stove lighter.}\label{msr}
\end{figure}

``Piezoelectricity'' is the term for the electric potential generated by a mechanical deformation of a particular type of crystal. Of interest in high-voltage applications is the fact that, for most crystals, there is a linear regime over which the output voltage is proportional to the applied stress. A device which rapidly deforms a robust crystal, in many cases by striking the crystal with a spring-loaded hammer, can produce a potential of many kV. One such device, the MSR camp-stove lighter of Figure~\ref{msr}, is a particularly robust model, typically providing $>10^5$ ``clicks'' without failure of the crystal or striking mechanism.\footnote{This number is empirical, but approximate. During the testing for HiCal-2, there was never a failure of a device under test, including units which had been used for multiple duration studies, equaling well over $10^5$ clicks.} In this case, the impact of the hammer with the crystal results in an HV discharge across the $\sim$6~mm space between the core and the sheath of the protruding metallic tube. 

The benefits of using such a device as an HV source are two-fold. One, the power required to generate the HV discharge is solely that needed to depress the clicker mechanically, which can be done using a motor and camshaft, as described below, which for HiCal-2 was measured to be 140~mW RMS. Second, the HV source itself is electrically isolated from the electromechanical operation of the ``clicker'', which is advantageous in a radio frequency application.

\subsection{A Model for high-voltage Discharge}

The ideal HiCal pulse would resemble a delta function in time, thereby probing the Antarctic surface with equal power at every frequency. To approach this ideal, HiCal-1 implemented a standard spark-gap transmitter, as used in late 19th century radio experimentation and telegraphy. Spark-gap transmitters 
consist of an antenna attached either in series or in parallel to an air-gap and a high-voltage source. The high voltage source generates a potential across the air gap such that an arc occurs between its nodes\cite{raizer}. During breakdown, RF is produced and transmitted via the antenna. Spark-gap transmitters are attractive for our purposes because the resultant RF is generally highly broadband, due to its transient nature.  
HiCal-2 improved upon the design of the spark-gap transmitter by using a finely-tunable spark gap hub to tune the gap itself in such a way as to maximize output amplitude and minimize pulse width.

\begin{figure}[H]
\centering
\includegraphics[width=.6\textwidth]{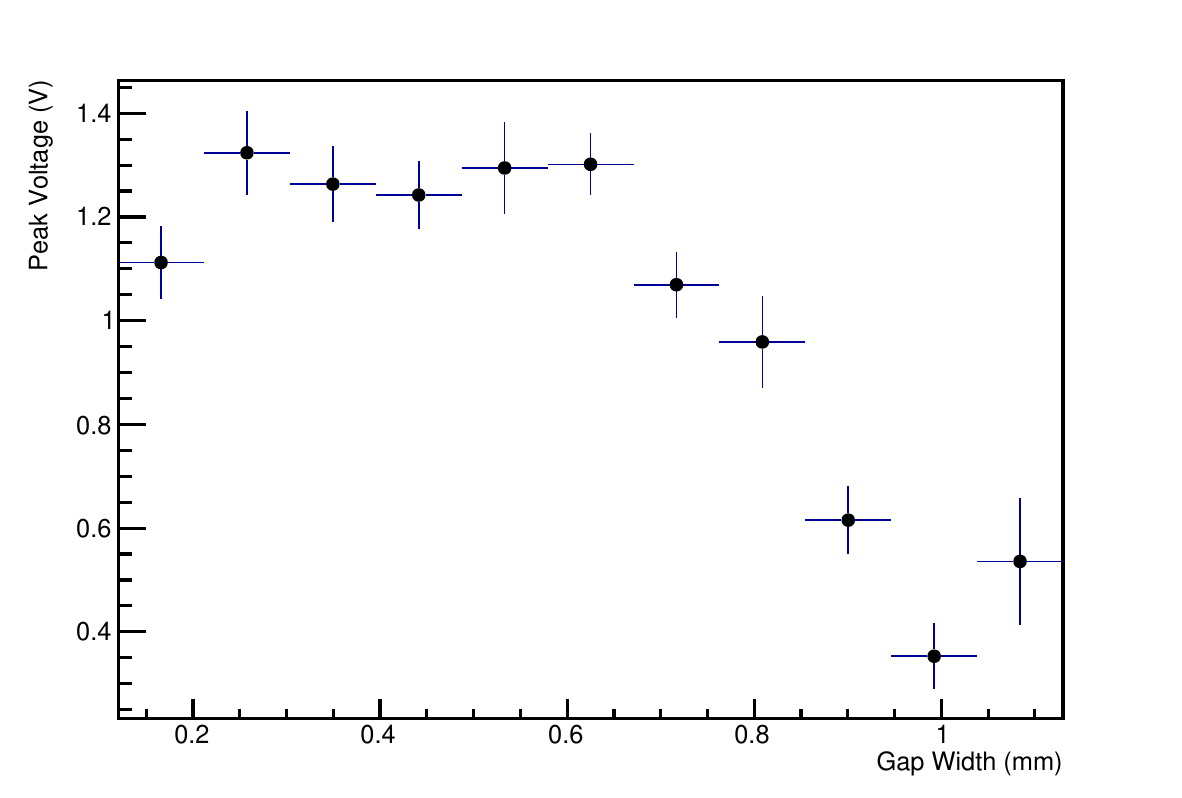}
\caption{Width of a spark gap vs. amplitude of the resultant pulse as measured into an RF antenna at 20m.}\label{gap_compare}
\end{figure}

Empirically, we observed an inverse relationship between spark gap size and RF emission amplitude for spark gap lengths shorter than 2~mm, as shown in Figure~\ref{gap_compare}.

\section{Instrument}

\subsection{high-voltage System}

\begin{figure}[H]
\centering
\includegraphics[width=.8\textwidth]{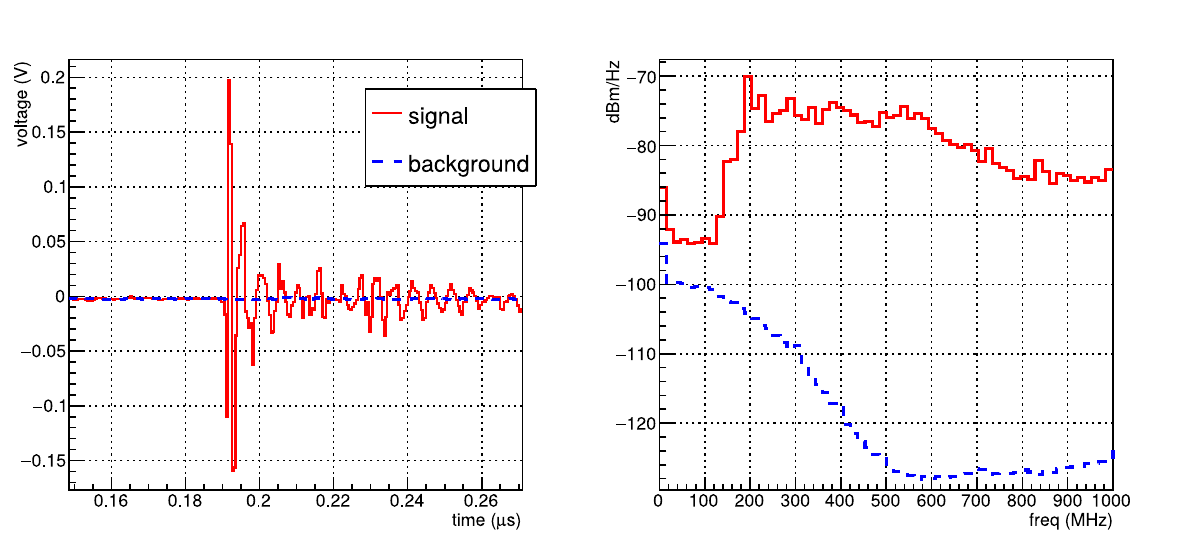}
\caption{An example pulse from the HiCal-2 system, as received by an Anita-4 horn antenna at a distance of 40~m and read out by an HP54542 oscilloscope, compared to ambient background. Left: Voltage vs. time. Right: Power spectrum.}\label{pulse}
\end{figure}

\begin{figure}[H]
\centering
\includegraphics[width=.5\textwidth]{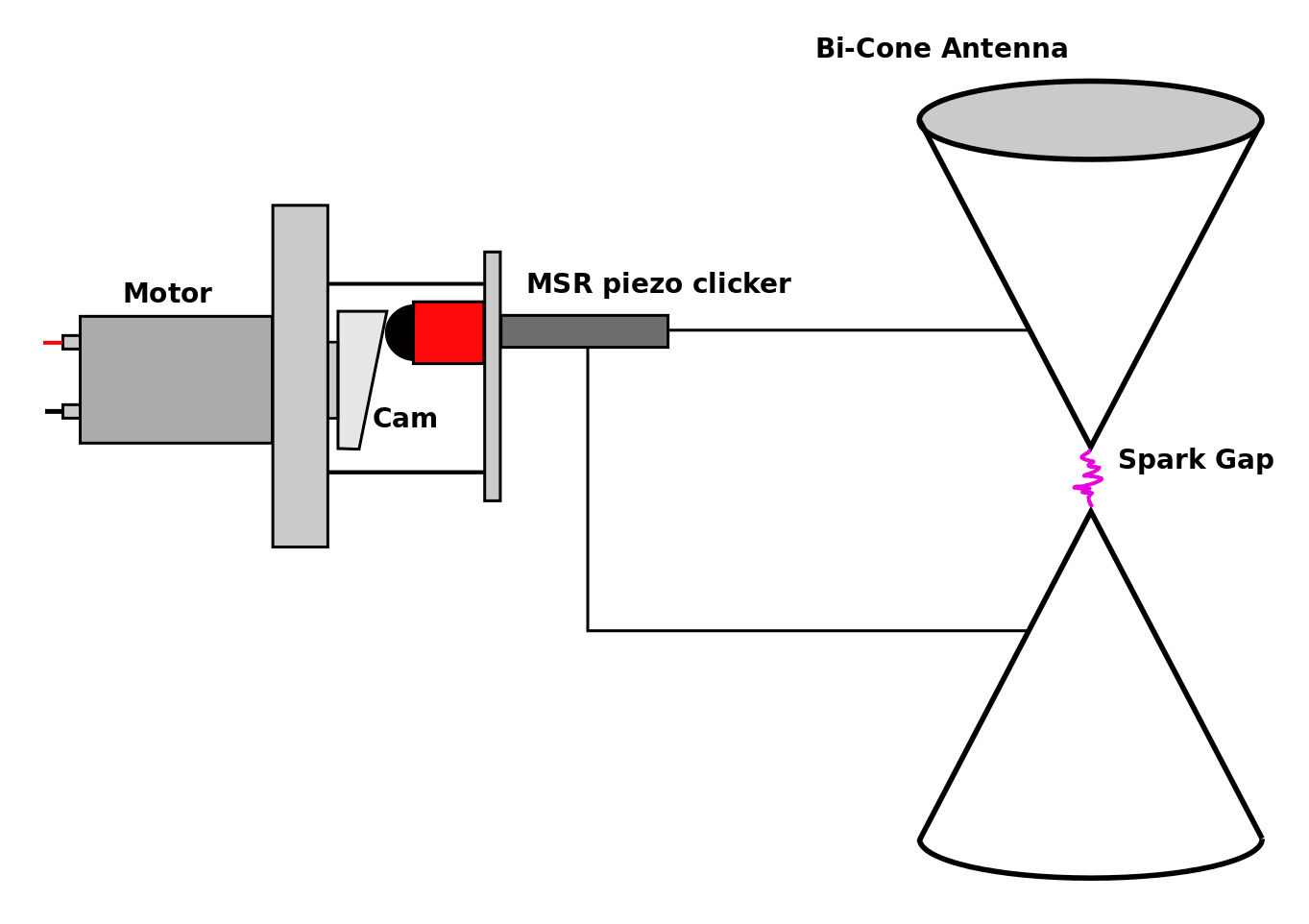}
\caption{Schematic of the motor mount, MSR piezo unit, and the bicone antenna (not to scale). The spark gap is exaggerated for illustration.}\label{hical_actual}
\end{figure}

HiCal-2 implements the MSR sparker as the HV source of Figure~\ref{hical_actual}. A servo motor (ServoCity 20~RPM gear motor) turns a cam (a rotor with a well-defined pitch angle) at a rate of $\sim $0.1~Hz that depresses the spring of the MSR sparker, thus generating an HV potential once per revolution. This sparker is connected to the antenna such that half of the bicone is attached to the core of the sparker, the other half to the sparker's metallic sheath. As long as the spark gap between the antenna halves is the minimum impedance path in the circuit, the breakdown location migrates from the MSR sparker core/sheath to the antenna. This breakdown causes an arc across the gap between the two halves of the bicone antenna, as shown in Figure~\ref{hical_actual}.

A key benefit of this system is that the primary failure mode is well-defined. Because the HV generation is decoupled from the electronics that turn the motor, any electronics failure will simply result in no pulses, as opposed to an unpredictable discharge.

\begin{figure}[H]
\centering
\includegraphics[width=.5\textwidth]{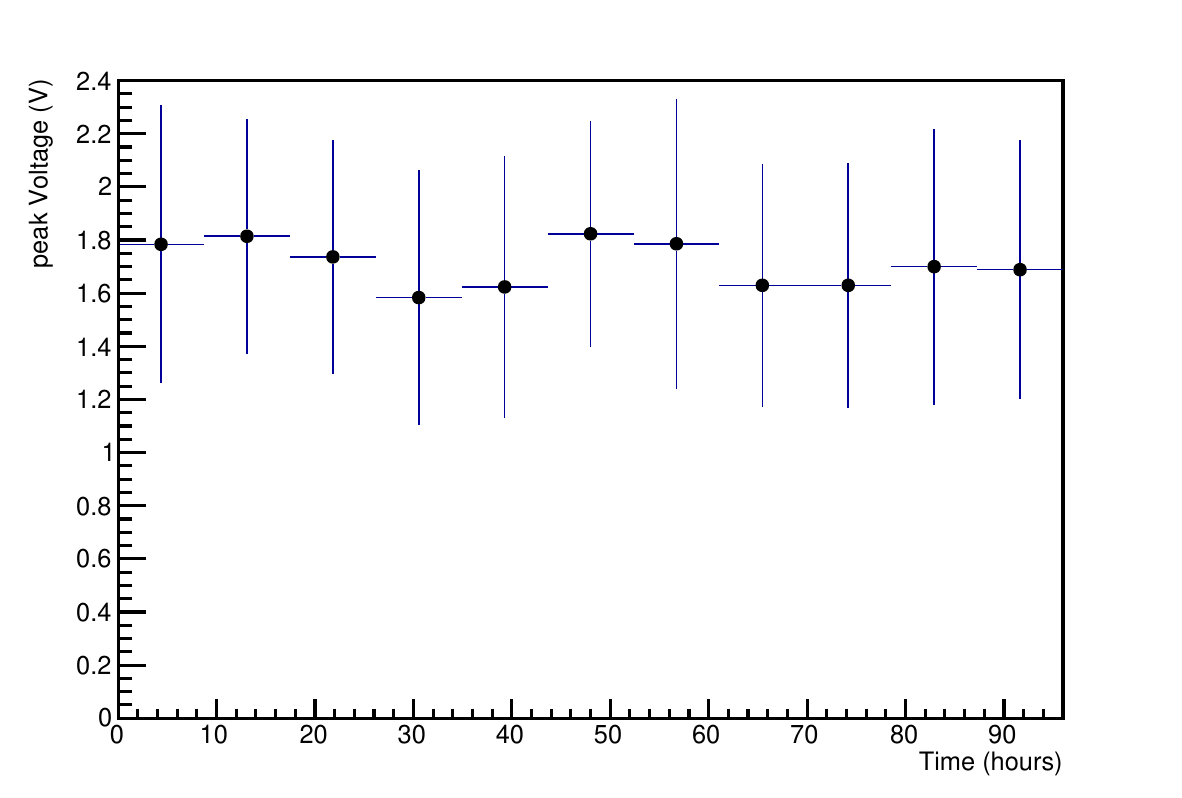}
\caption{A 96-hour spark test, showing the consistency of the RF emission. Error bars indicate the full spread of recorded values, with a standard deviation in peak voltage of 0.13~V.}\label{duration}
\end{figure}

\begin{figure}[H]
\centering
\includegraphics[width=.8\textwidth]{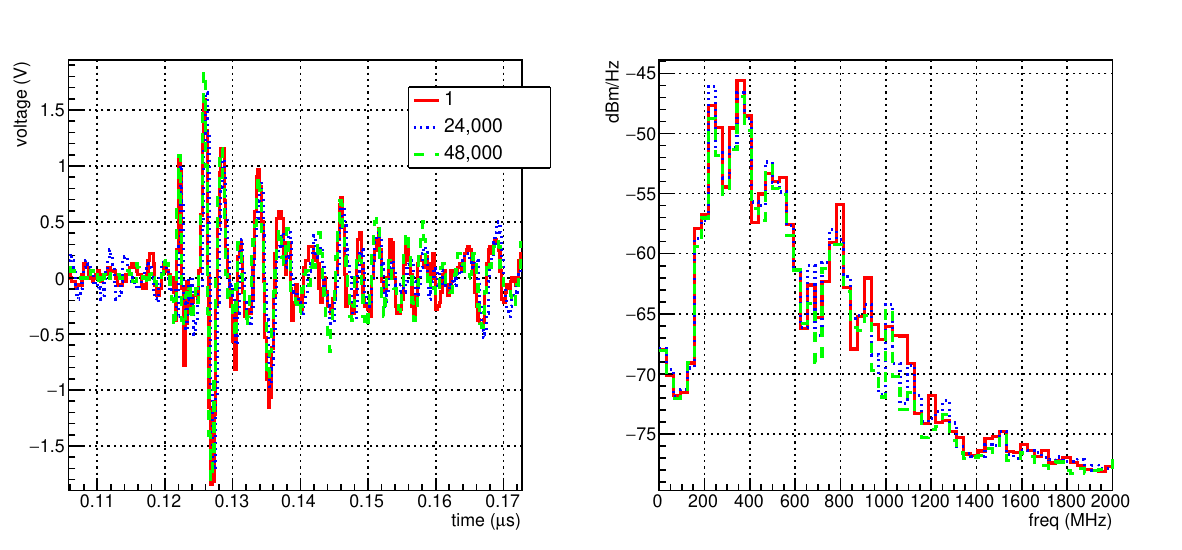}
\caption{Comparisson of several pulses from a 90+ hour duration test, showing consistency over time. Left:time. Right: Power spectrum.}\label{consistent}
\end{figure}

The MSR clicker was chosen for its durability. Several duration tests were performed to ensure that the device would continue to perform after tens of thousands of clicks, and in various environments. For one such test, the motor was run continuously for 96 hours, and the output pulses were recorded. Figure~\ref{duration} shows the peak received amplitude as a function of time during the test. Figure~\ref{consistent} shows a comparison of the first, the 24,000th, and 48,000th pulse from the 96 hour duration test. These pulses look nearly identical in both time and frequency space, demonstrating the durability of the piezo/breakdown system.

There was some concern about the breakdown changing the chemistry of the atmosphere inside of the sealed PV such that the breakdown and subsequent RF emission would change over the course of the mission. During the long test described above, there was no observation of any significant change in the pulses from start to finish, and therefore it was concluded that such an effect was not noticeable in the volume of air within the PV for the number of pulses expected during the HiCal-2 mission.

\subsection{Antenna}
\begin{figure}[H]
\centering
\begin{minipage}[t]{.4\linewidth}
\centering
  \includegraphics[width=.8\textwidth]{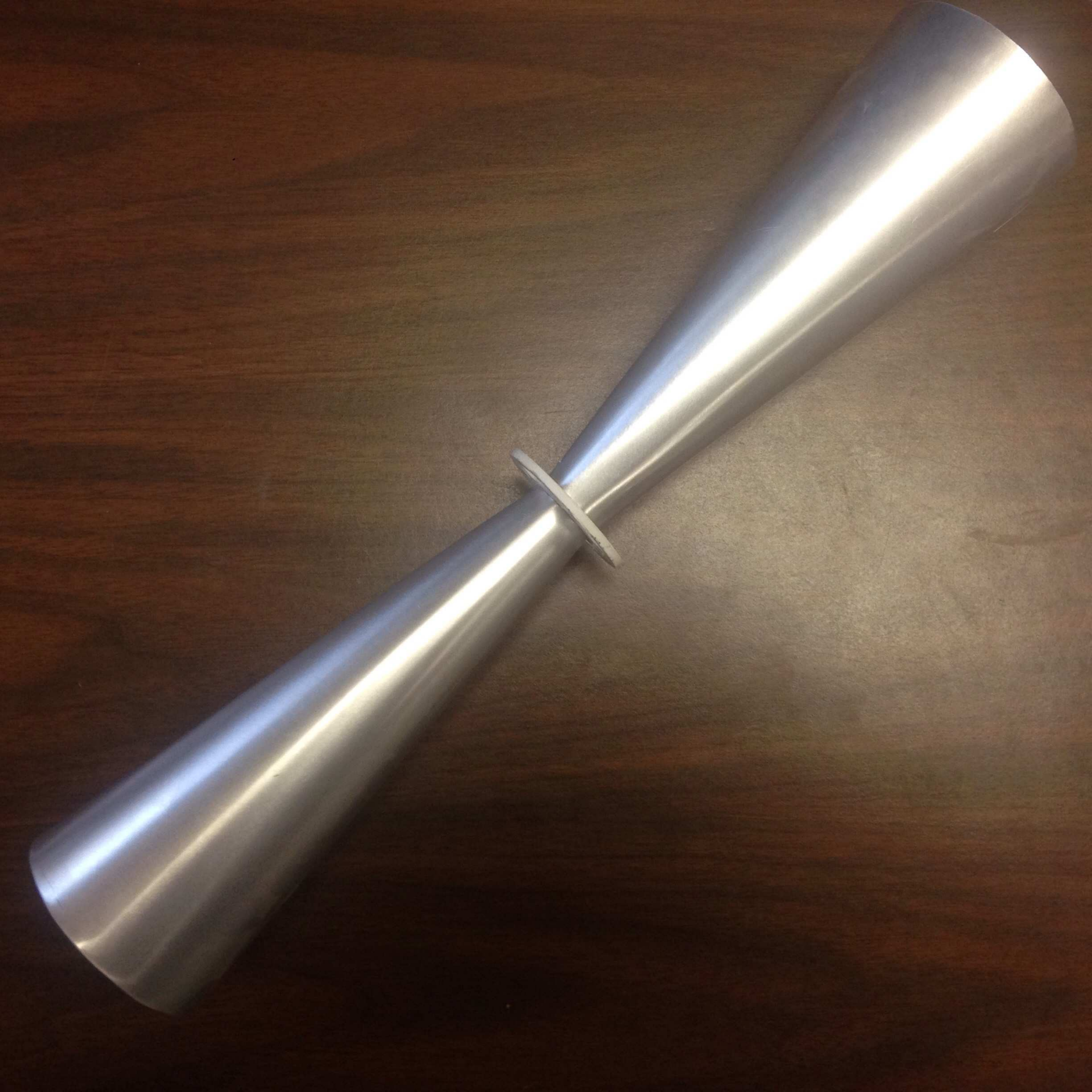}

\end{minipage}
\qquad
\begin{minipage}[t]{.4\linewidth}
\centering
  \includegraphics[width=.8\textwidth]{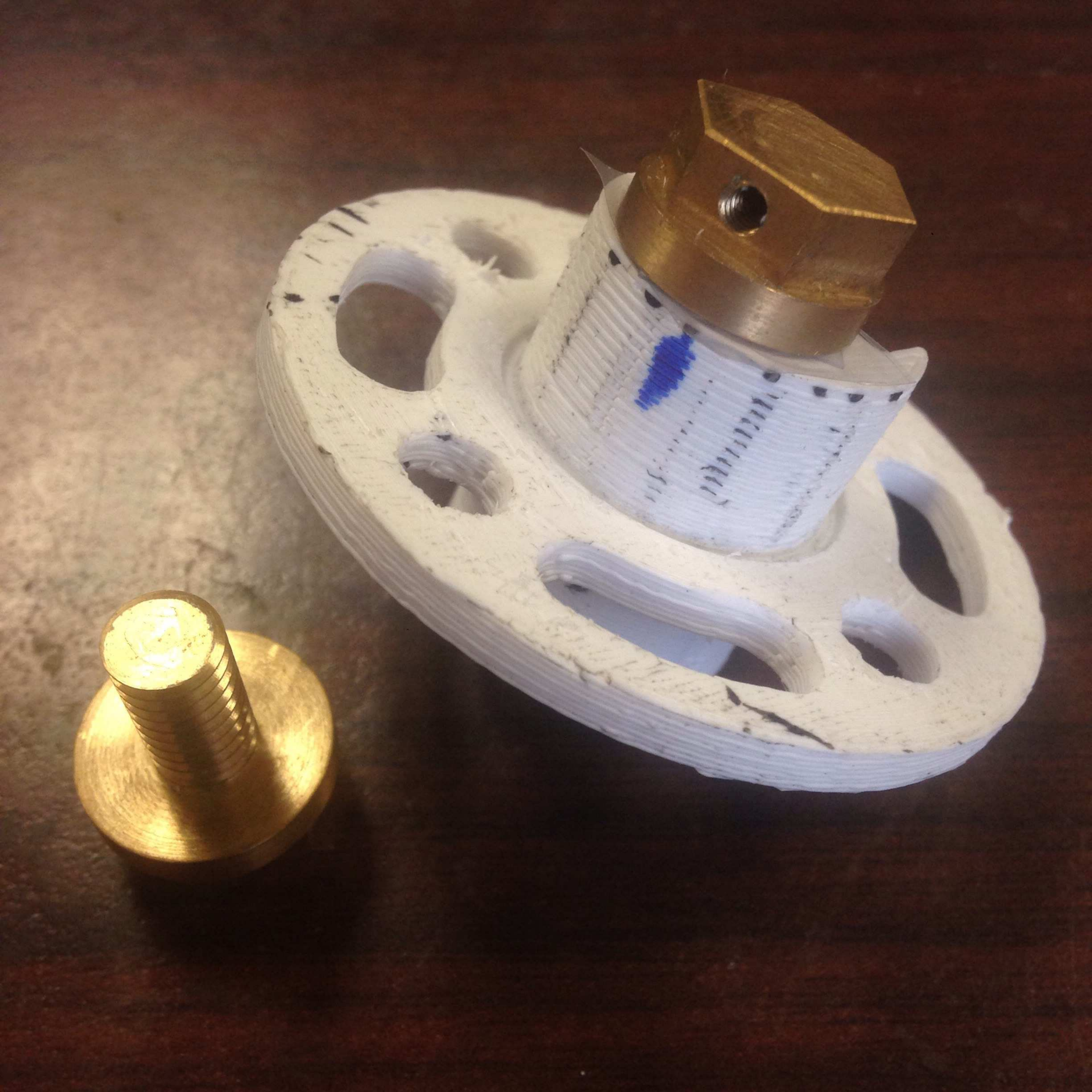}
\end{minipage}
\caption{The HiCal-2 bicone antenna, left, with a detail of the 3d-printed spacer and machined spark gap set bolt, right.}\label{bicone}
\end{figure}

The HiCal-2 bicone antenna of Figure~\ref{bicone} was used produce the RF emission from the controlled HV discharge. This antenna needed to be broadband in the range of ANITA-4 sensitivity (200--1200~MHz) but sufficiently small and light to satisfy the mass constraints for flight. The antenna consists of 24~gauge (.51~mm) aluminum sheet cut and rolled into the bicone configuration, with a half-length of 20~cm and a radius of 4.5~cm. A machined spacer, shown in Figure~\ref{bicone}, separates the two antenna halves, isolating them from one another. Threaded into this spacer is a tunable spark gap, set to roughly 230$\mu$m. The spacer is coupled to the antenna via brass set screws. The antenna's rigidity is provided by the pressure vessel itself, with an inner dimension matching the bicone outer dimension for a snug fit. 

\begin{figure}[H]
\centering
\includegraphics[width=.5\textwidth]{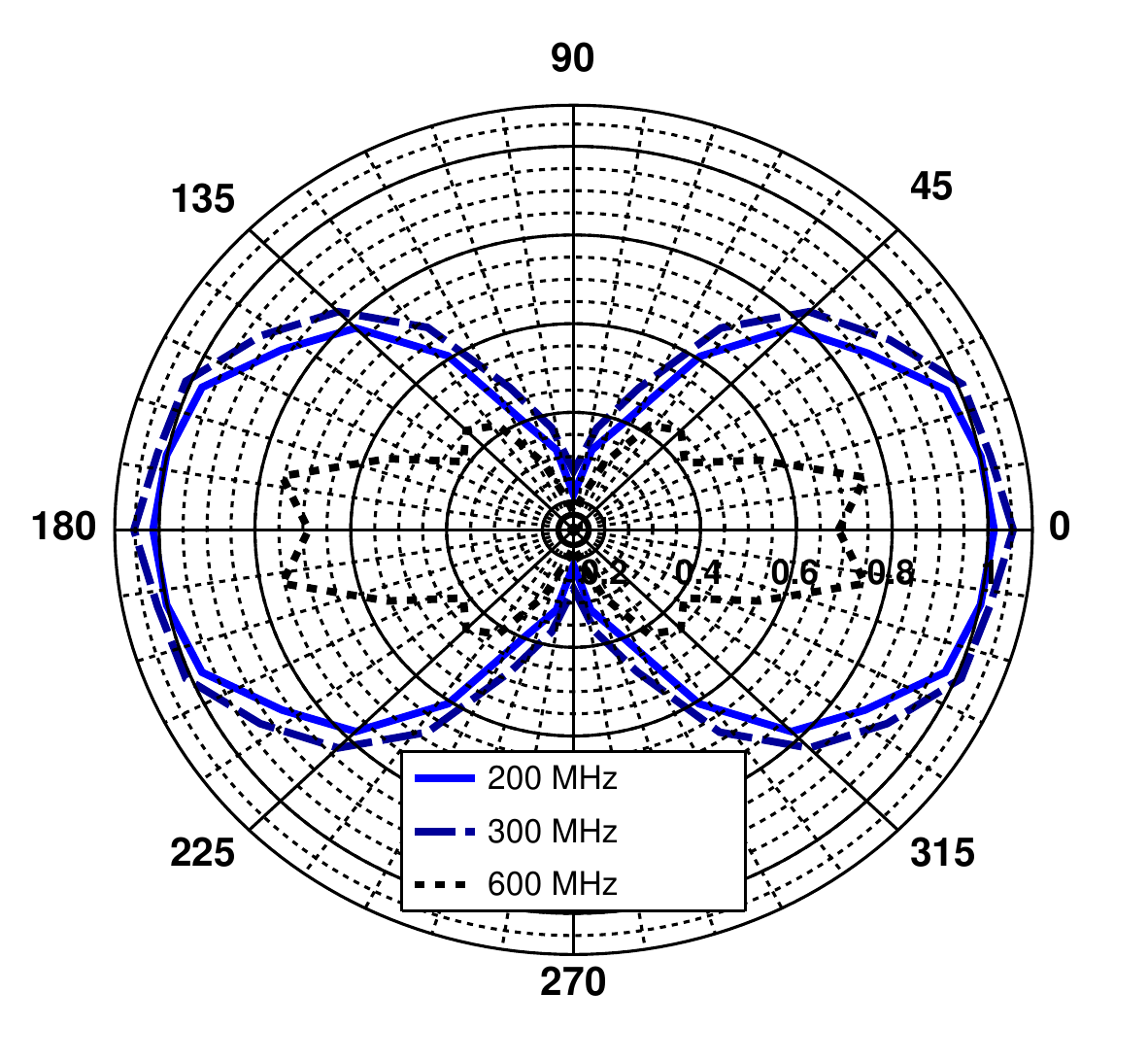}
\caption{The HiCal-2 antenna beam pattern for 3 frequencies. The antenna axis is aligned along the vertical axis of the plot, with cylindrical symmetry assumed.}
\label{pattern}
\end{figure}

The antenna flies in the horizontal polarization configuration, e.g. aligned so that the axis of cylindrical symmetry of the antenna flies parallel to the ground  plane. This ensures that the gain of the antenna in the direction of ANITA-4 is the same as the gain in the direction of the ice below (except for the extremes near the bicone ends). The gain pattern of the antenna is very similar to a dipole antenna, although more broadband due to the bicone design. The pattern for several different frequencies is given in Figure~\ref{pattern}.

\subsection{Pressure Vessel and Timestamp System}
\begin{figure}[H]
\centering
\includegraphics[width=.5\textwidth]{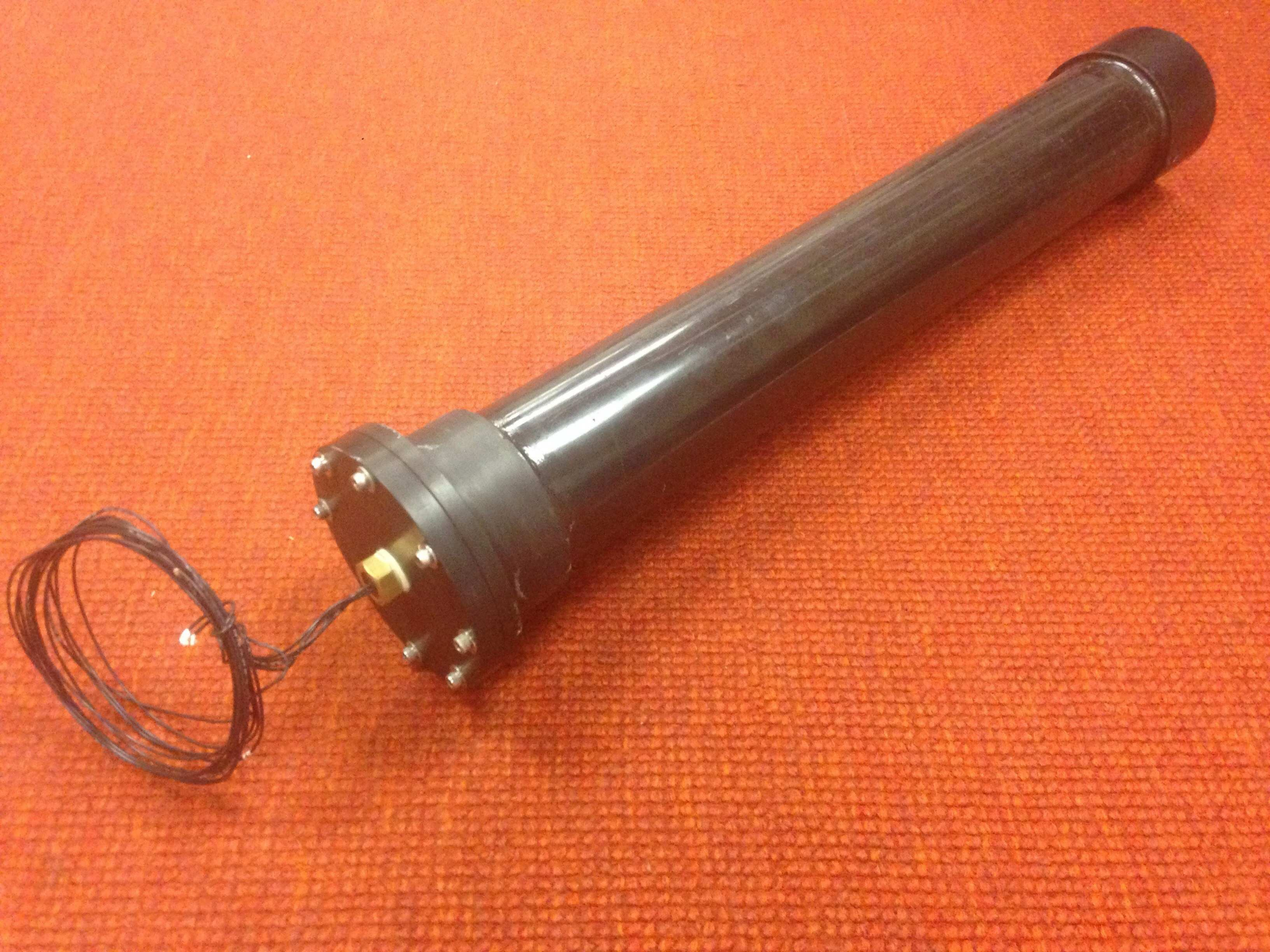}
\caption{The HiCal-2 pressure vessel. 9 cm in diameter by 60 cm in length.}
\label{pv}
\end{figure}

\begin{figure}[H]
\centering
\includegraphics[width=.5\textwidth]{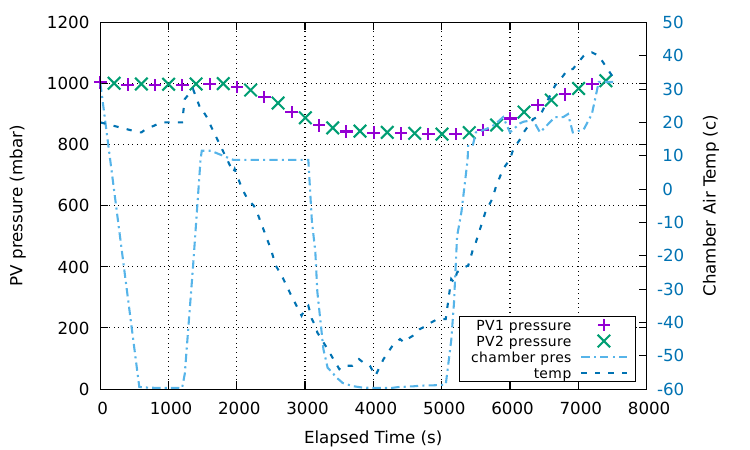}
\caption{Performance of the pressure vessel in a thermal/vacuum test inside of a thermal/vacuum chamber, for which temperature and pressure were cycled through flight conditions. Note the dip in pressure due to the cold temperatures typical of ascent through the troposphere.}
\label{bemco}
\end{figure}

To maintain the integrity and consistency of the breakdown, the antenna environment must be maintained near 1000~mbar within the ambient environment during flight of 5~mbar. Therefore, the HV system and antenna were housed in a 1 atm pressure vessel (PV). 
Weight and dielectric requirements motivated the use of a lightweight plastic. Acrylonitrile butadiene styrene (ABS), a typical plastic piping used for drainage, was used due to its affordability and light weight. The pressure vessel consisted of a main body with a 7.6~cm interior diameter terminated on one end by a fixed cap and the other end terminated by a machined ABS flange. An end cap with an o-ring was bolted to this flange to provide the seal. The end cap was fitted with an epoxy-filled threaded feedthrough, into which were fixed 4 wires, allowing for power and data transfer into and out of the vessel. The vessel is shown in Figure~\ref{pv}, and the performance of the pressure vessels during a rigorous thermal/vacuum test at NASA's Columbia Scientific Ballooning Facility (CSBF) in Palestine, TX is shown in Figure~\ref{bemco}.

\begin{figure}[H]
\centering
\includegraphics[width=.5\textwidth]{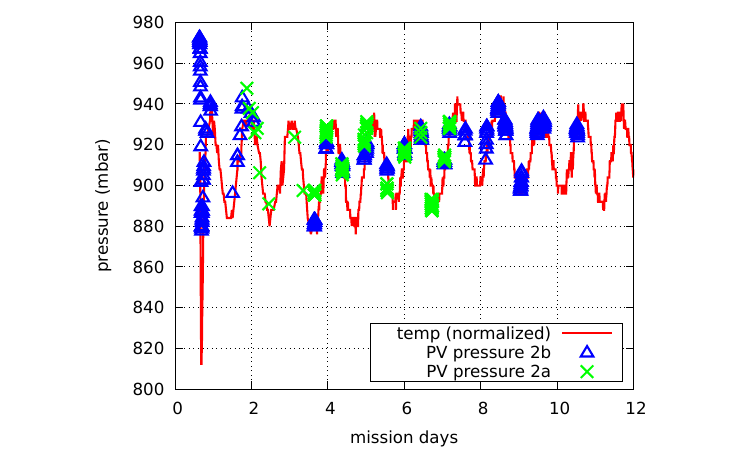}
\caption{Pressure of the HiCal-2 PVs. For reference, the ambient temperature is overlaid, showing that the pressure tracks the general trend of the ambient temperature. The intermittent nature of the pressure vessel data is due to the operation of HiCal in flight. The instruments were activated for roughly 30 minutes, 3 times per day; pressure monitors are only powered at that time. }
\label{pvflight}
\end{figure}

The antenna and HV system resided within the PV. Because the HiCal motor runs autonomously when powered and is not triggered or tied to a GPS clock, the pulse emission time is only predictable to within the $\sim$10~ms jitter in the camshaft rotation period, necessitating passive timestamping of the output pulses. To this end, a rigid 10~gauge (2.5~mm) pickup wire was threaded axially into the bicone, to detect the pulse and transmit it to the ATSA board (described below) for timestamping. The PV also housed a Honeywell SSCDANN015PAAB5 pressure monitor, to track the performance of the pressure vessel through the flight. The motor and the pressure monitor were both fed in parallel from a 5~V source. Correspondingly, the 4 feedthrough wires through the endcap flange of the PV were power (5V), ground, pressure monitoring, and pulse pickup antenna for timestamping. The time-series data of the monitors for HiCal-2a (second HiCal balloon to be launched) and 2b (first HiCal payload to launch) are shown in Figure~\ref{pvflight}, showing that the PV held pressure through the flight, and that it served as a de-facto thermometer, sensitive to the day/night cycle. 

\subsection{ATSA}
The azimuth and time-stamp apparatus (ATSA) flew on top of HiCal-2 to provide azimuthal orientation (accurate to $\pm$3 degrees) and RF pulse timestamp information (accurate to $\pm$55~$\mu$s)  for the transmitted pulses. Details of this device are given in \cite{atsa}.

\subsection{HiCal system board}

The HiCal system board featured a PIC microcontroller, and was responsible for formatting all of the science data for telemetery, and providing conditioned power to ATSA and the HV discharge system, including the PV pressure monitor. It included an internal clock, synced to GPS, that would latch when ATSA registered a pulse from the PV. This timestamp, along with temperature and azimuth, were formatted and sent via RS-232 to the NASA telemetery unit for each pulse.

\subsection{NASA electronics and power}
The NASA electronics, called the MIP~\cite{mip}, consisted of GPS and Iridium units, for position and telemetery purposes respectively. The instrumentation package on board HiCal-2 had a throughput of 255 bytes/minute, including `housekeeping' data, such as GPS, temperature, and PV pressure monitoring, and `science' data, including our timestamped pulses and azimuth information. The MIP also had commanding capability, which allowed us to remotely turn the system on and off through an electromechanical relay governing the power to the HiCal system board.

NASA also provided custom batteries for the instrument, designed specifically for the extreme conditions of flight. These batteries powered the MIP and HiCal electronics, via the aforementioned relay. Each payload had a set of two 9~V, 31 Amp-hour battery packs.

\section{Implementation and Testing}
The systems were designed and built at various locations, with the final integration performed over the summer of 2016 at CSBF.  At the end of September 2016, a `hang test' (Fig. \ref{launch}, left panel) was performed, during which HiCal GPS and communication were tested, followed by shipment to Antarctica in mid-October, 2016.

\section{Flight}
\begin{figure}[H]
\centering
\begin{minipage}[t]{.4\linewidth}
\centering
  \includegraphics[width=.5\textwidth]{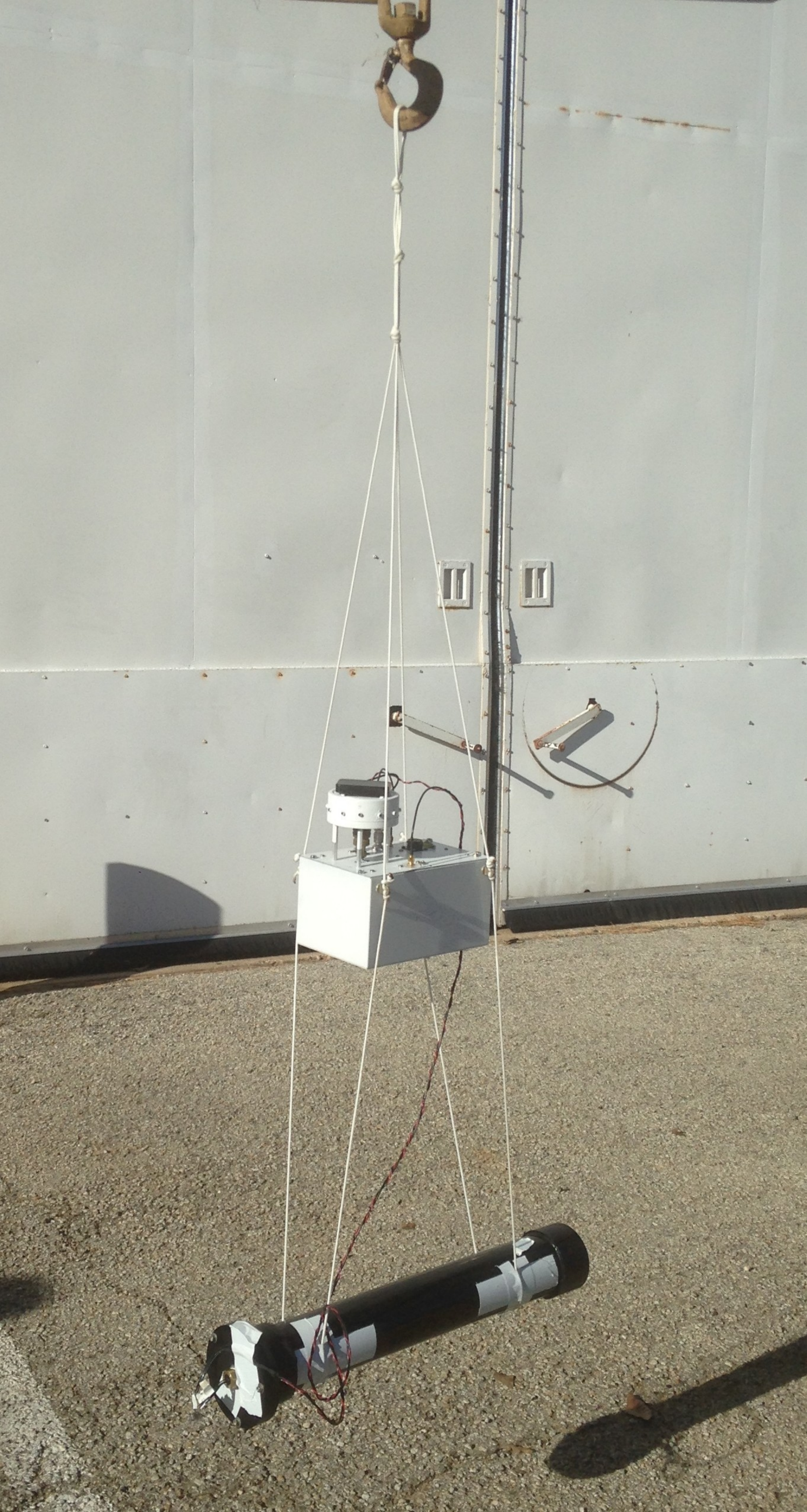}

\end{minipage}
\qquad
\begin{minipage}[t]{.4\linewidth}
\centering
  \includegraphics[width=.6\textwidth]{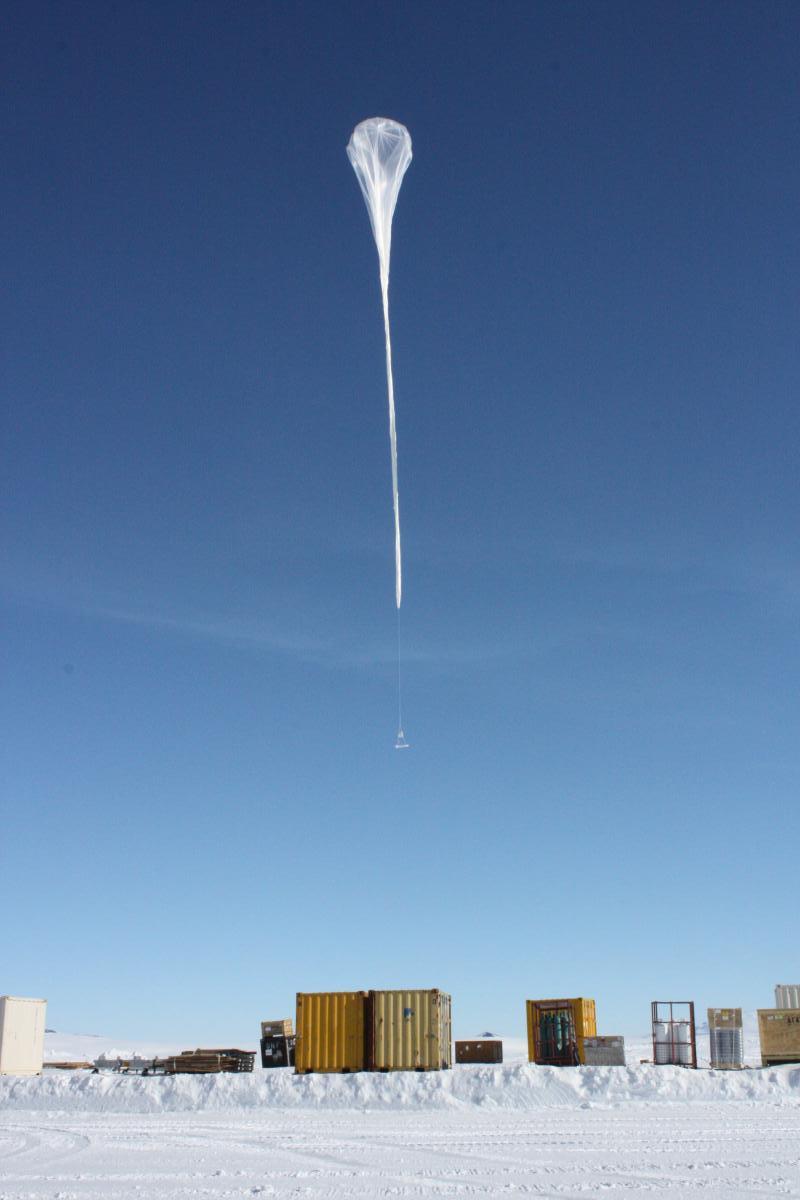}
\end{minipage}
\caption{Left: The hang test at CSBF, where all systems performed nominally, verifying payload-readiness for the Antarctic flight. Right: The HiCal-2a launch on the Ross Ice Shelf.}\label{launch}
\end{figure}

ANITA 4 launched on December 3 2016 from NASA's Long Duration Ballooning Facility (LDB) on the Ross Ice Shelf of Antarctica. Although the original plan was to launch HiCal immediately after ANITA 4, logistics and weather made that impossible, requiring a delay until ANITA swung back close enough to LDB after its first revolution to allow a launch in proximity. HiCal-2b launched  on December 11 (initially leading ANITA-4 by approximately 700 km), with HiCal-2a launching on December 12 (initially trailing ANITA-4 by approximately 500 km). A photo of HiCal-2a just after launch is shown in the right panel of Figure~\ref{launch}.  The flight paths of the HiCal payloads tracked well with the ANITA 4 instrument, ranging in line-of-sight distance to ANITA from 100-700km over the flight lifetime.

\begin{figure}[H]
\centering
\includegraphics[width=.5\textwidth]{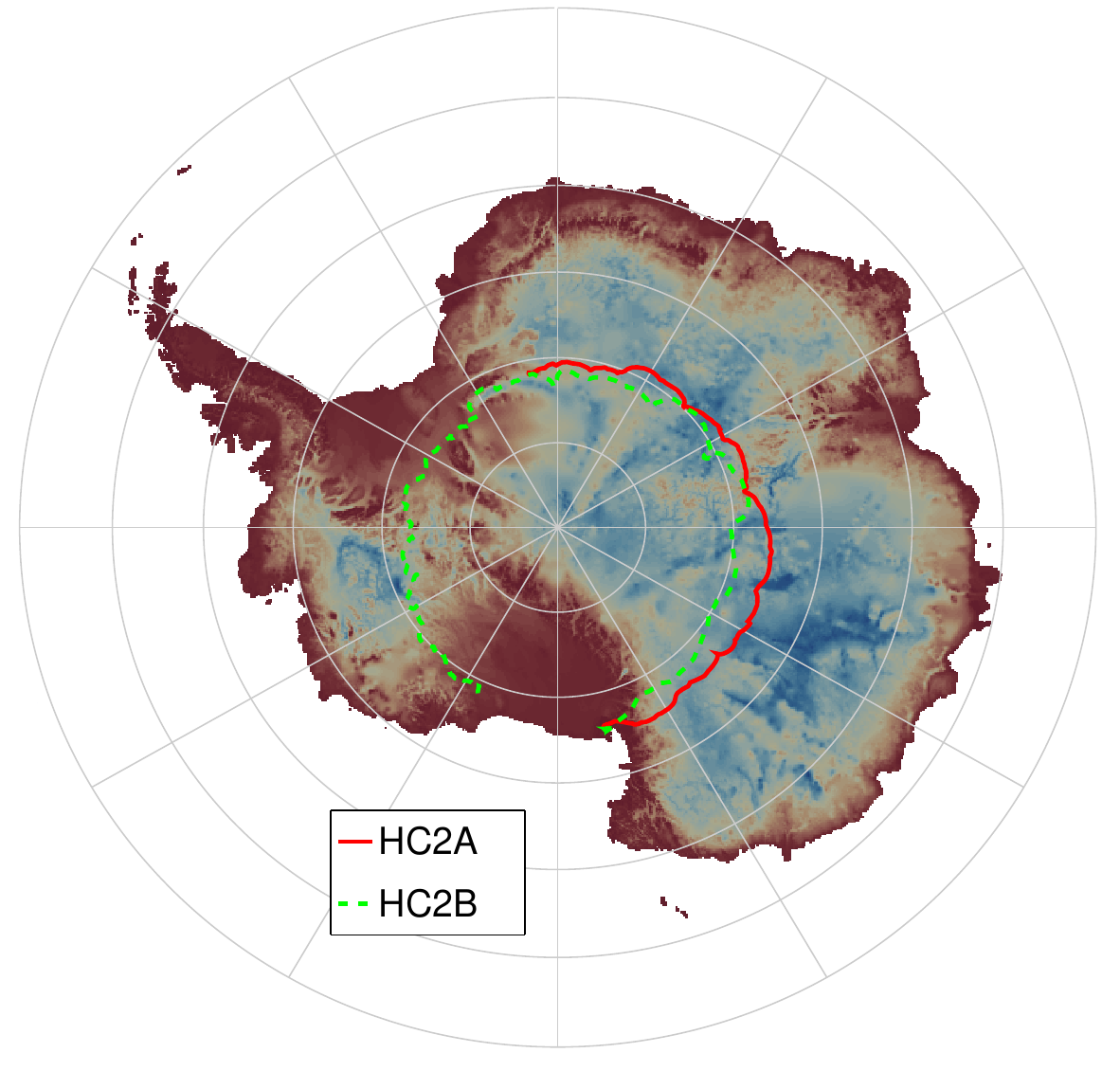}
\caption{The HiCal flight paths during the ANITA4 flight, over a map of the Antarctic continent. The color gradient indicates ice thickness\cite{bedmap}.}\label{flight}
\end{figure}

To extend the lifetime of the instruments and thereby sample the largest possible spread of Antarctic terrain, HiCal-2a and HiCal-2b were turned on to pulse in 30~minute intervals approximately once every six hours, and left in a low power state the rest of the flight. Although both instruments performed well, one instrument died at one-half its expected lifetime -- this may have been due to a failure in one of the 2 onboard battery packs, although the exact failure remains undiagnosed as the payloads were not recovered. Together the two payloads completed 1.5 cumulative revolutions of the continent for a combined 18 days of flight. Their flight paths are shown in Figure~\ref{flight}.

\section{Summary and Discussion}
The analysis of the HiCal-2 data is underway, and will be presented more completely in a companion article. ANITA-4 captured ${\cal O}$(10,000) pulses total, direct and reflected from the ice surface, from HiCal-2, with over 2,500 `pairs' analyzed thus far, for which both the direct and reflected impulses from the same pulse were recorded. This dataset provides surface roughness information at observation angles (measured relative to the horizontal) of 4--30 degrees (separation distances of 100-700~km), and has allowed a high-statistics measurement of the surface roughness coefficient of the Antarctic ice. This distance/angle range covers the full range of interest for surface roughness studies. The range of distances (and therefore angles) scanned by HiCal-2 overlaps with all previous surface reflectivity measurements mentioned in this article.

HiCal-3, which is scheduled to fly with the proposed ANITA-5 instrument, will feature a larger payload, both horizontal and vertical polarizations, higher telemetry throughput, local RF pulse capture and improved time-stamping. This will allow us to have a record of the emitted pulse as well as the pulse as captured by ANITA, which will lower the systematic uncertainty in pulse-to-pulse variation during analysis. 

\section{Acknowledgments}
This work was supported by NASA grant number NNX15AC20G and the U.S. National Science Foundation Office of Polar Programs. The authors would like to thank CSBF's engineers and launch team for their excellent mission support, A. Hase and the U. of Kansas machine shop, and J. Roth of U. of Delaware.

\bibliography{/home/natas/Documents/physics/tex/bib}{}

\end{document}